# Physics Educators as Designers of Simulation using Easy Java Simulation ( Ejs )

Loo Kang WEE

Ministry of Education, Education Technology Division, Singapore
wee_loo_kang@moe.gov.sg, weelookang@gmail.com

Abstract: To deepen the professional practice of physics educators, I seek to highlight the Open Source Physics (OSP) and Easy Java Simulation (Ejs or EJS) community of educators that engage, enable and empower teachers as learners so that we can be leaders in our teaching practice. I learned through Web 2 online collaborative means to develop simulations together with reputable physicists through the open source digital library. By examining the open source codes of the simulation through the Ejs toolkit, I was able to examine and make sense of the physics from the computational models created by practicing physicists. I will share some of the simulations that I have remixed from existing library of simulations models into suitable learning environments for inquiry of physics. http://www.phy.ntnu.edu.tw/ntnujava/index.php?board=28.0
.
Keyword: easy java simulation, active learning, education, teacher professional development, e-learning, applet, design, open source, GCE Advance Level physics
PACS: 01.50.H-, 07.05.Tp, 01.50.Lc, 83.10.Rs

## I. INTRODUCTION

Easy Java Simulations (Ejs) is a software tool (java code generator) designed for the creation of discrete computer simulations.

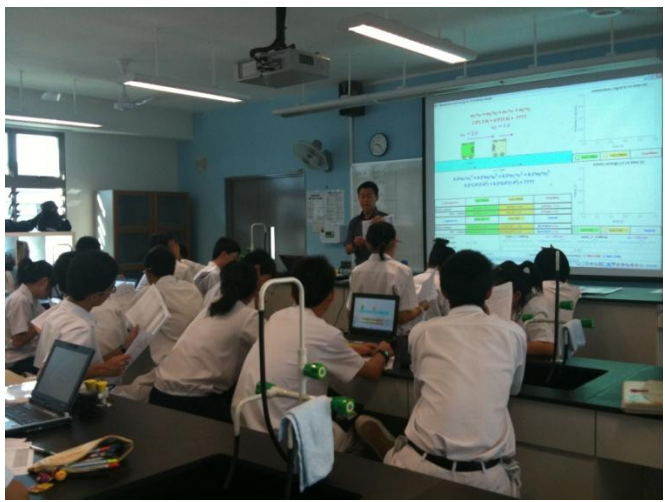

Figure 1. A typical physics laboratory with a simulation as learning environment. The photo shows the teacher giving the students a brief introduction to the guided inquiry tasks and affordances of the simulation.

I have created serveral computer models, also known as simulation, to allow our students to visualize Physics phenomena, using a free authoring toolkit called Easy Java Simulation [1].

Building on open source codes shared by the Open Source Physics (OSP) community and with help from Fu-Kwun's NTNUJAVA Virtual Physics Laboratory [2], I have customized serveral computer models as tools to support inquiry learning, downloadable from digital libraries in NTNUJAVA Virtual Physics Laboratory, creative commons attribution licensed.

Interested readers could refer to this Youtube [3] and blogpost [4] of the actual session.

## II. HOW CAN EJS BE USED?

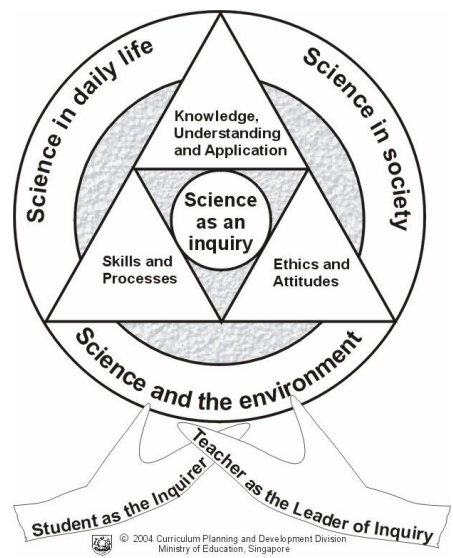

Figure 2. The Ministry of Education Singapore Science Curriculum Framework [5]

My experiences in creating simulations for inquiry pedagogy enactment suggests teachers could create these virtual laboratory simulations to deepen their own professional practice through informal learning-networking with the global Open Source Phsyics community. Students can conduct inquiry learning of science (Figure 2) while more ambitious tasks include creating their own simulations as representations of their understanding.

## III. SEVEN COMPUTER MODELS

To add to the body of knowledge of teacher created-remixed simulations, I would briefly highlight some of my contributions to the simulation design, allowing the readers to judge for themselves whether these simulations are useful tools for active learning purposes. Anyone is welcome to change the source codes of the simulations to suit their own needs, licensed creative commons attribution.

*A. Simple Harmonic Motion Model*

This is one of my earliest remix where I change the user interface and color scheme. My contributions include the block and 2 spring system for setting the context of the spring mass system, vector arrows to represent quantities I normally drawn on pen paper problem and 7 scientific graphs (Figure 3) covered in the Physics syllabus.

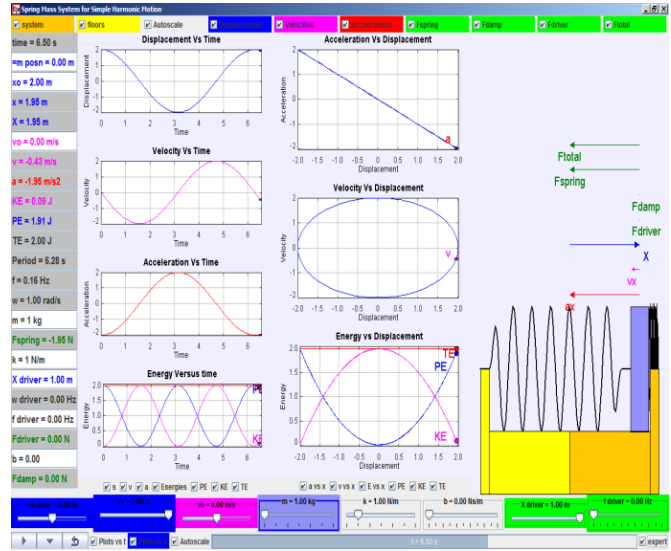

Figure 3. Simple Harmonic Motion Model [6] derived from Francisco's original work found in Ejs examples customized to show the 7 scientific representations and a driving force system commonly taught in the A level Physics.

*B. Collision Carts Model*

This is my second remix where my contributions include mathematical representations (Figure 4) to illicit predictive thinking about the concepts, table of data and scientific graphs for inquiry data collection etc.

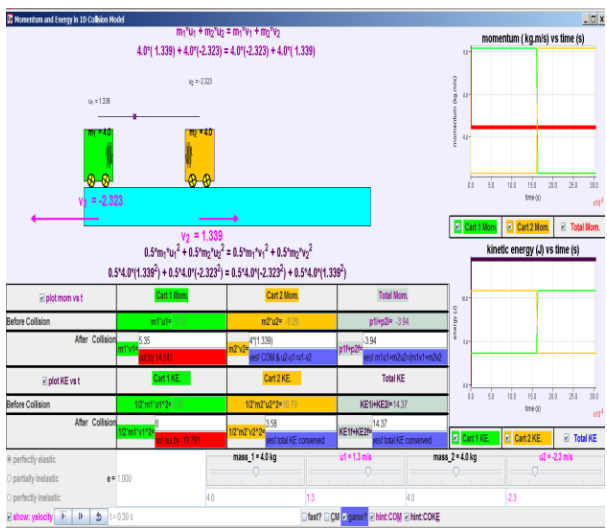

Figure 4. Collision carts (ideal) model [7] derived from Francisco's original work [8] showing mathematical representations to illicit predictive thinking about the concepts.

*C. Geostationary Orbit Model*

This model (Figure 5) is a result of a new feature in EJS using new Java 3D implementation as unveiled during Multimedia in Physics Teaching and Learning Conference MPTL 14 [9]. This model though not as suited for inquiry, can do used as a visualization tool with 3 different geostationary orbits and many other non-geostationary orbits to serve as counter-examples to deepen the concept. My contributions include the satellite, different geostationary and non-geostationary orbits etc.

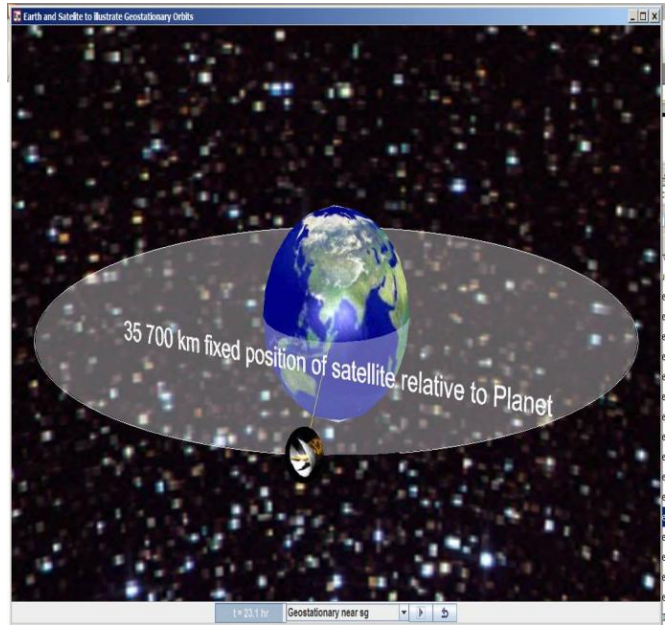

Figure 5. Geostationary orbit model [10] derived from Francisco's original work [11] showing a geostationary orbit with different options of geostationary and non-geostationary orbits.

*D. Circular Motion on a table Model*

This 3D model (Figure 6) is by Fu-Kwun Hwang which allows student to explore the horizontal circular motion provided only the radius is varied for perfect circular motion. My contributions include the forces visualization and variables that can be used for inquiry.

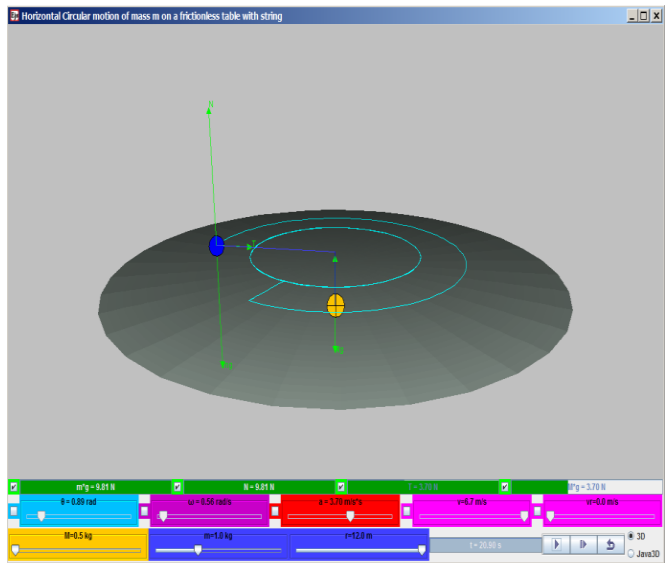

Figure 6. Circular Motion on a table Model [12] derived from Fu-Kwun's original work [13] showing a 3D view of a table and 2 masses attached together with a string connected through a hole at the centre of the table. This allows for visualization of the forces on the masses as the blue mass moves on the table in horizontal circular motion

### E. Alternating Current Generator Model

This 3D model (Figure 7) is of my most complex remix where my contributions include using another Electric Generator Model [14] by Wolfgang Christian & Anne Cox to create the galvanometer and I made many of the variables to enable inquiry such as area of coil, external magnetic field, number of turns of the coil for investigative inquiry.

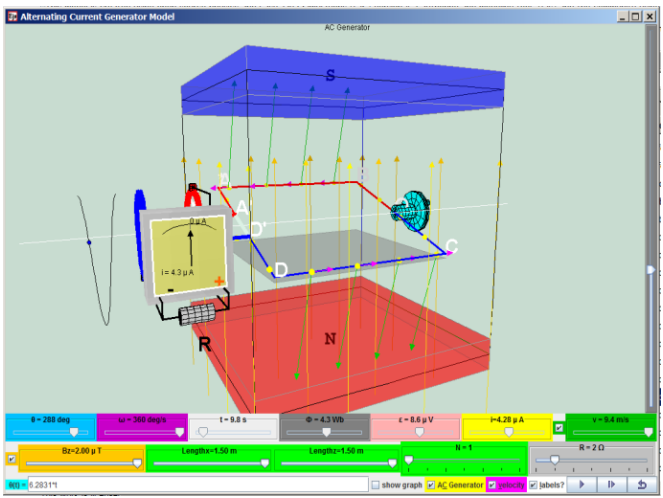

Figure 7. Alternating Current Generator Model [15] derived from Fu-Kwun's original work [16] that is a 3D model with variables such as area of coil, external magnetic field, number of turns of the coil for investigative inquiry.

### F. Vernier Calipers Model

This model (Figure 8) is a result of remixing from many models [17-19] into a single customized model suited for Singapore syllabus. My contirbutions include adding hints, input field of assessment of learning and different versions of the vernier scale to suit different makes of real vernier calipers.

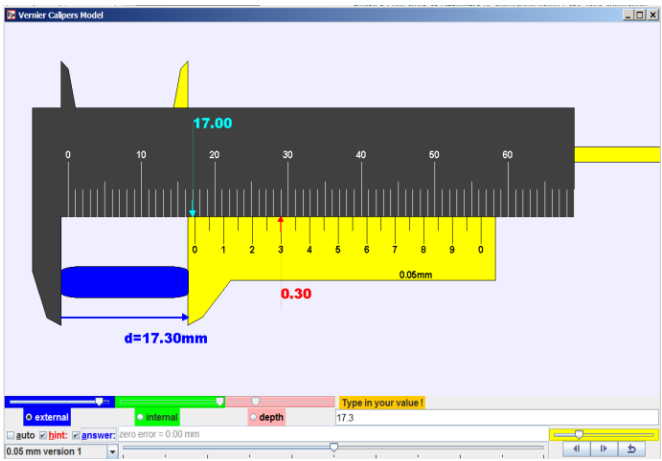

Figure 8. Vernier Calipers Model [20] derived from Fu-Kwun's original work [17] with hints and answers with possibility of different vernier scale and an input field for students to test their own understanding.

### G. Bar Magnet & Compass Model

This model (Figure 9) is similar to a PhET Interactive Simulations Magnet and Compass [21] where my contributions include the earth semi transparent graphics and minor edits to the codes in the compass. While changing the source codes, I realise how much I have learnt about computational physics as well as physics concepts modelled.

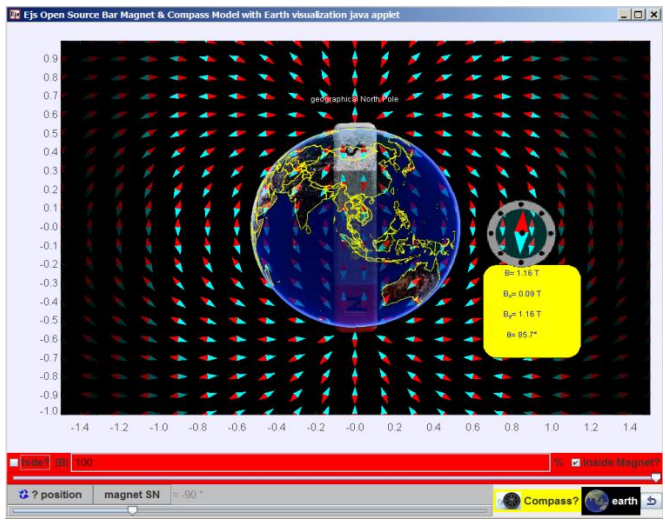

Figure 9. Bar Magnet & Compass Model [22] derived from Christian, Francisco and Anne's original work [23] showing a bar magnet and its relationship to Earth's magnetic field and a compass with numerical data.

## IV. CONCLUSION

I have shared some of my remixed simulations as an indication of how other teachers may deepen their professional practice in physics education, through informal professional learning with the Open Source Physics (OSP) and Easy Java Simulation (EJS) community.

I shared briefly some of the simulations that I have remixed from existing library of simulations models into suitable learning environments for inquiry of physics.

I hope more teachers will find the simulation useful in their own classes and join in the journey of remixing simulations as a form of self directed professional learning to deepen our own teaching practices.

### ACKNOWLEDGEMENT

I wish to acknowledge the passionate contributions of Francisco Esquembre, Fu-Kwun Hwang and Wolfgang Christian M. Belloni, A. Cox, W. Junkin, H. Gould, D. Brown, J. Tobochnik, Jose Sanchez, J. M. Aguirregabiria, S. Tuleja, M. Gallis, T. Timberlake, A. Duffy, T. Mzoughi, and many more in the EJS and OSP community for their simulations ideas and curriculum activities.

### REFERENCE

[1]  F. Esquembre. (2012, 13 September). *Easy Java Simulations*. Available: http://www.um.es/fem/EjsWiki/pmwiki.php

[2]  F.-K. Hwang. (2010, 13 September). *NTNU Virtual Physics Laboratory*. Available: http://www.phy.ntnu.edu.tw/ntnujava/index.php & http://www.phy.ntnu.edu.tw/ntnujava/index.php?board=23.0

[3] L. K. Wee, "Physics Educators as Designers of Simulation using Easy Java Simulation ( Ejs ) ", ed. Singapore: YouTube https://www.youtube.com/watch?v=Ek09SYaAzfQ, 2010.

[4] L. K. Wee. (2010). *AAPT 2010 Conference Presentation:Physics Educators as Designers of Simulations*. Available: http://www.aapt.org/Conferences/sm2010/loader.cfm?csModule=security/getfile&pageid=25474
https://docs.google.com/present/edit?id=0ATIvSg-TzZrZZGZmN3dnYzdfMjEzZmh3dnZzZ2c&hl=en & http://weelookang.blogspot.com/2010/07/physics-educators-as-designers-of.html

[5] MOE. (2008). *Science Syllabus Primary 2008*. Available: http://www.moe.gov.sg/education/syllabuses/sciences/files/science-primary-2008.pdf

[6] L. K. Wee and F. Esquembre. (2008). *Ejs open source simple harmonic motion java applet SHM virtual lab*. Available: http://www.phy.ntnu.edu.tw/ntnujava/index.php?topic=758.0

[7] L. K. Wee and F. Esquembre. (2008). *Ejs open source java applet 1D collision carts Elastic and Inelastic Collision* [application/java]. Available: http://www.phy.ntnu.edu.tw/ntnujava/index.php?topic=831.0

[8] F. Esquembre. (2009). *Collision in one dimension* [application/java]. Available: http://www.um.es/fem/EjsWiki/Main/ExamplesCollision1D

[9] C. Eick, L. Meadows, and R. Balkcom, "Breaking into Inquiry: Scaffolding Supports Beginning Efforts to Implement Inquiry in the Classroom," *Science Teacher,* vol. 72, p. 5, 2005.

[10] L. K. Wee and F. Esquembre. (2010). *Ejs Open Source Geostationary Satellite around Earth Java Applet (requires Java 3D and Runtime ed.)*. Available: https://sites.google.com/site/lookang/edulabgravityyearthandsatelliteyjc/ejs_EarthAndSatelite.jar?attredirects=0&d=1 & http://www.phy.ntnu.edu.tw/ntnujava/index.php?topic=1877.0 (requires Registration to download)

[11] F. Esquembre. (2010). *Ejs Open Source Earth and Moon Model (1.0 ed.)*. Available: http://www.phy.ntnu.edu.tw/ntnujava/index.php?topic=1830.0

[12] F.-K. Hwang and L. K. Wee. (2010). *Ejs Open Source Horizontal Circular Motion of Mass on a table java applet* [application/java]. Available: http://www.phy.ntnu.edu.tw/ntnujava/index.php?topic=1883.0

[13] F.-K. Hwang. (2010). *Conservation of Angular momentum and 3D circular motion* [application/java]. Available: http://www.phy.ntnu.edu.tw/ntnujava/index.php?topic=1454.0

[14] A. Cox and W. Christian. (2009). *Electric Generator Model (1.0 ed.)*. Available: http://www.compadre.org/Repository/document/ServeFile.cfm?ID=9218&DocID=1236

[15] F.-K. Hwang and L. K. Wee. (2009). *Ejs Open Source Alternating Current Generator Model Java Applet ( AC Generator ) (December 29, 2010 ed.)*. Available: http://www.phy.ntnu.edu.tw/ntnujava/index.php?topic=1940.0

[16] F.-K. Hwang. (2009). *A loop rotating in a magnetic field (How electric power generator works!) (December 29, 2010 ed.)*. Available: http://www.phy.ntnu.edu.tw/ntnujava/index.php?topic=915.0

[17] F.-K. Hwang. (2007). *Vernier Caliper and Micrometer with zero error options*. Available: http://www.phy.ntnu.edu.tw/ntnujava/index.php?topic=567.0

[18] F.-K. Hwang. (2010). *vernier cliper (inch/mm mode, dial mode, measure inner/outer diameter and depth)*. Available: http://www.phy.ntnu.edu.tw/ntnujava/index.php?topic=567.0

[19] F.-K. Hwang and L. K. Wee. (2009). *Vernier Caliper Model*. Available: http://www.compadre.org/Repository/document/ServeFile.cfm?ID=9707&DocID=1445

[20] F.-K. Hwang and L. K. Wee. (2012). *Ejs open source Vernier calipers java applet with objects, help & 0-error logic*. Available: http://www.phy.ntnu.edu.tw/ntnujava/index.php?topic=684.0

[21] M. Dubson, C. Malley, K. Perkins, C. Wieman, D. Harlow, and A. Paulson. (2011). *PhET: Magnet and Compass (Version: 2.07 Build Date: (Oct 7, 2011) ed.)*. Available: http://phet.colorado.edu/en/simulation/magnet-and-compass

[22] W. Christian, F. Esquembre, A. Cox, and L. K. Wee. (2009). *Ejs Open Source Bar Magnet & Compass Model with Earth visualization java applet*. Available: http://www.phy.ntnu.edu.tw/ntnujava/index.php?topic=1210.0

[23] W. Christian, F. Esquembre, and A. Cox. (2009). *Magnetic Bar Field Model (1.0 ed.)*. Available: http://www.compadre.org/Repository/document/ServeFile.cfm?ID=9414&DocID=1309

## AUTHOR

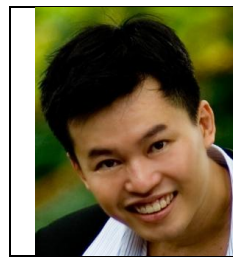

Loo Kang WEE is currently an educational technology specialist at the Ministry of Education, Singapore. He was a junior college physics lecturer and his research interest is in Open Source Physics tools like Easy Java Simulation for designing computer models and use of Tracker.